\documentclass[superscriptaddress,twocolumn,showpacs,amsmath,amssymb]{revtex4}

\usepackage{graphicx}

\begin{document}

\title{
Test of finite temperature RPA on a Lipkin model}

\author{K. Hagino}
\affiliation{ 
Department of Physics, Tohoku University, Sendai 980-8578,  Japan} 

\author{F. Minato}
\affiliation{
Innovative Nuclear Science Research Group, 
Japan Atomic Energy Agency, Tokai, Ibaraki 319-1195, Japan}


\begin{abstract}
We investigate the applicability of finite temperature random 
phase approximation (RPA) using a solvable Lipkin model. 
We show that the finite temperature RPA reproduces reasonably well 
the temperature dependence of total strength, both for the positive 
energy (i.e., the excitation) and the negative energy (i.e., 
the de-excitation) parts.  
This is the case even at very low temperatures, which may be 
relevant to astrophysical purposes. 
\end{abstract}


\maketitle

The random phase approximation (RPA) \cite{RS80} and its extension, 
quasi-particle RPA (QRPA), have successfully described nuclear giant 
resonances\cite{LG76,GK02}. 
Usually, those giant resonances are built on the ground state, but 
the collective states can be excited also from 
excited states \cite{B55}. 
In the early 80's, heavy-ion fusion experiments revealed 
the existence of giant resonances in hot nuclei
\cite{N81,B89,S86,Y90,K99}, 
and properties of collective excitation at finite temperatures 
have attracted much interest. 
To discuss giant resonances in hot nuclei, 
RPA has been extended 
by including thermal effects 
\cite{VVM84,FER83,RREF84,ER93,Som83,SB84,BRT84,CBD84}. 
Such extension is referred to as finite temperature RPA or thermal RPA. 

Recently, there have been renewed interests 
in atomic nuclei at finite temperatures, 
in connection to nuclear astrophysics\cite{RBR99,LKD01,NK04,CHMR00,NPVM09}. 
See also Ref. \cite{KGG04} for collective excitations in hot exotic nuclei. 
The finite temperature RPA has often been used to estimate {\it e.g.}, 
beta decay rates in a stellar environment \cite{LKD01,NK04,CR99}.
In order to make quantitative calculations for nuclear astrophysical 
purposes, especially for r-process nucleosynthesis, it is 
necessary to know the accuracy of finite temperature RPA also at relatively low 
temperatures. 

The aim of this paper is to assess the applicability of finite temperature 
RPA using the Lipkin-Meshkov-Glick model \cite{LMG65}. This 
is a schematic solvable model and has been employed extensively 
to test many-body methods \cite{RS80}. 
A similar study for finite temperature RPA 
has already been done by Rossignoli and Ring \cite{RR98} 
(see also Ref. \cite{TEKO94}), but they explicitly investigated 
only the positive energy 
part of a strength function. 
Here we investigate both the positive and negative energy parts 
separately. 
Since the sensitivity to the thermal occupation probability is 
large for deexcitation process at low temperatures, such investigation 
provides an interesting test of finite temperature RPA. 
We also consider both the canonical and the 
grand canonical ensembles for the 
exact solutions of the Lipkin model, while 
Ref. \cite{RR98} considered only the grand canonical ensemble. 
This is important because the finite temperature RPA is based on the 
grand canonical ensemble despite that the number of particle 
is well-defined in actual nuclei. 
For compound nuclei formed in heavy-ion fusion 
reactions, the grand canonical ensemble may be justified because of 
neutron evaporation processes\cite{Som83}. However,  
it is not obvious whether the same argument holds for nuclei in 
a stellar condition at low temperatures. 
By comparing the results with canonical ensemble 
to those with grand canonical ensemble, one can get some insight 
about the applicability of many-body theories based on the grand canonical 
ensemble, such as finite temperature RPA. 

In the Lipkin model, one considers two single-particle levels, at 
energy of $-\epsilon/2$ and $\epsilon/2$, respectively, each of 
which has 2$\Omega$-fold degeneracy. 
The Hamiltonian for this model reads\cite{LMG65}
\begin{equation}
H=\epsilon\hat{K}_0-\frac{V}{2}(\hat{K}_+\hat{K}_++,\hat{K}_-\hat{K}_-),
\end{equation}
where $V$ is the strength of a two-body interaction. 
The operators $\hat{K}_0, \hat{K}_+$ and $\hat{K}_-$ are defined as,
\begin{eqnarray}
\hat{K}_0&=&\frac{1}{2}\,\sum_{i=1}^{2\Omega}
(c_{1i}^\dagger c_{1i}-c_{0i}^\dagger c_{0i})\\
\hat{K}_+&=&\sum_{i=1}^{2\Omega}c_{1i}^\dagger c_{0i},~~~\hat{K}_-=(\hat{K}_+)^\dagger.
\end{eqnarray}
Here, $c_{0i}^\dagger$ and $c^\dagger_{1i}$ are the creation operators for the 
lower and upper levels, respectively. 

The exact solutions of the Lipkin model can be obtained with the quasi-spin 
formalism \cite{RS80,LMG65}. The eigen states are then classified 
in terms of the eigen value of the operator $\hat{K}^2=\hat{K_0}^2
+(\hat{K}_+\hat{K}_++\hat{K}_-\hat{K}_-)/2$. 
Denoting those states and their energy as $|J\alpha\rangle$ and 
$E_{J\alpha}$, respectively, the strength function for 
the canonical ensemble is given by \cite{RR98}, 
\begin{eqnarray}
S_C(E)&=&\frac{1}{Z_C}\sum_{J,\alpha,\alpha'}Y_C(J)\,e^{-\beta E_{J\alpha}}
|\langle J\alpha'|\hat{F}|J\alpha\rangle|^2 \nonumber \\
&&\times\delta(E-E_{J\alpha'}+E_{J\alpha}),
\label{sc}
\end{eqnarray}
where $\beta=1/kT$ is the inverse temperature and $\hat{F}$ is 
the transition operator. We have assumed that $\hat{F}$ does not change 
the value of $J$. $Y_C(J)$ is the degeneracy of the $J$ state given 
by \cite{RPM91} 
\begin{eqnarray}
Y_C(J)&=&W_C(J)-W_C(J+1)(1-\delta_{J,J_{\rm max}}), \\
W_C(J)&=&
\left(
\begin{array}{c}
2\Omega \\
N/2-J
\end{array}
\right)
\left(
\begin{array}{c}
2\Omega \\
N/2+J
\end{array}
\right),
\end{eqnarray}
where $N$ is the number of particle in the system, and 
$J_{\rm max}={\rm min}[N/2,2\Omega-N/2]$ is the maximum $J$ for given $N$. 
$Z_C$ in Eq. (\ref{sc}) is the partition function given by 
\begin{equation}
Z_C=\sum_{J,\alpha}Y_C(J)e^{-\beta E_{J\alpha}}.
\end{equation}

In this paper, we consider only a system with $N=2\Omega$ 
(that is, half-filling). In this case, the chemical potential $\mu$ in 
the grand canonical ensemble is zero, and the exact strength function 
for the grand canonical ensemble is given by a similar formula as in 
Eq. (\ref{sc}) but with a different value of degeneracy \cite{RPM91}, 
\begin{eqnarray}
Y_{GC}(J)&=&W_{GC}(J)-W_{GC}(J+1)(1-\delta_{J,J_{\rm max}}), \\
W_{GC}(J)&=&
\left(
\begin{array}{c}
4\Omega \\
2\Omega-2J
\end{array}
\right).
\end{eqnarray}

In addition to the exact solution, 
we also seek an approximate solution for the strength function 
using the finite temperature RPA. To this end, 
we first solve the thermal Hartree-Fock equation\cite{RS80,VVM84,TEKO94}
\begin{equation}
h_{\rm HF}
\left(
\begin{array}{c}
D_{0k} \\
D_{1k}
\end{array}
\right)
=e_k
\left(
\begin{array}{c}
D_{0k} \\
D_{1k}
\end{array}
\right),
\end{equation}
with 
\begin{eqnarray}
(h_{\rm HF})_{00}&=&-\epsilon/2, ~~~
(h_{\rm HF})_{11}=\epsilon/2, \\
\\
(h_{\rm HF})_{01}&=&
(h_{\rm HF})_{01}=
-V(N-1)\sum_{k=0,1}f_kD_{0k}^*D_{1k},
\end{eqnarray}
where 
\begin{equation}
f_k=\frac{1}{1+e^{(e_k-\mu)/kT}},
\end{equation}
is the thermal occupation probability of the Hartree-Fock state $k$. 
With the Hartree-Fock basis, we assume that the excitation operator 
of the system is given by 
\begin{equation}
\hat{Q}^\dagger = x_{10}\sum_{i=1}^{N}a_{1i}^\dagger a_{0i}
+x_{01}\sum_{i=1}^{N}a_{0i}^\dagger a_{1i},
\end{equation}
where $a_{1i}^\dagger$ and $a_{0i}^\dagger$ are the creation operators for the 
Hartree-Fock states. 
The finite temperature RPA equation then reads \cite{VVM84},
\begin{equation}
\left(
\begin{array}{cc}
a&b \\
-b&-a
\end{array}
\right)
\left(
\begin{array}{c}
x_{10} \\
x_{01}
\end{array}
\right)
=
\omega
\left(
\begin{array}{c}
x_{10} \\
x_{01}
\end{array}
\right),
\end{equation}
with
\begin{eqnarray}
a&=&e_1-e_0+(f_0-f_1)\cdot \frac{1}{2}(N-1)V\sin^22\alpha, \\
b&=&-(f_0-f_1)\cdot (N-1)V\left(1-\frac{1}{2}\sin^22\alpha\right), 
\end{eqnarray}
where $D_{00}=\cos\alpha$. 
With the solutions of the RPA equations, the RPA strength function 
is obtained as \cite{VVM84,RREF84,ER93,CVVM90},
\begin{eqnarray}
S_{\rm RPA}(E)&=&\sum_n\frac{\omega_n}{|\omega_n|}\cdot 
\frac{1}{1-e^{-\beta E}}\,
\left|\sum_{kl}
\langle k|\hat{F}|l\rangle(f_k-f_l)x_{kl}^{(n)}\right|^2 \nonumber \\
&&\times\delta(\omega_n-E),
\end{eqnarray}
where the matrix elements for $\hat{F}$ are taken with the Hartree-Fock basis. 

Let us now solve the model Hamiltonian numerically and compute the 
strength function. 
For this purpose, we take the particle number to be $N=20$, and set 
$VN/\epsilon=0.5$. 
As a transition operator, we consider $\hat{F}=(\hat{K}_++\hat{K}_-)/2$. 
Following Ref. \cite{RR98}, we smear the strength function with a 
width of $\eta/\epsilon=0.1$. 

Figure 1(a) shows the strength function at temperature of $T/\epsilon$=0.05. 
For our choice of parameters, the difference in the strength function 
between the 
canonical and the grand canonical ensembles is small, and we only plot 
the result of 
grand canonical ensemble as the exact solution (see the solid line). 
The dashed and the dotted lines are the result of finite temperature RPA 
and thermal 
Hartree-Fock, respectively. At this low temperature, the thermal effect 
is almost negligible, and the 
strength function is actually almost the same as that at zero temperature. 
Notice that the RPA well reproduces the exact strength function. 
The Hartree-Fock result is not satisfactory, and 
the RPA correlation plays an important role. 

\begin{figure}[htb]
\includegraphics[clip,scale=0.5]{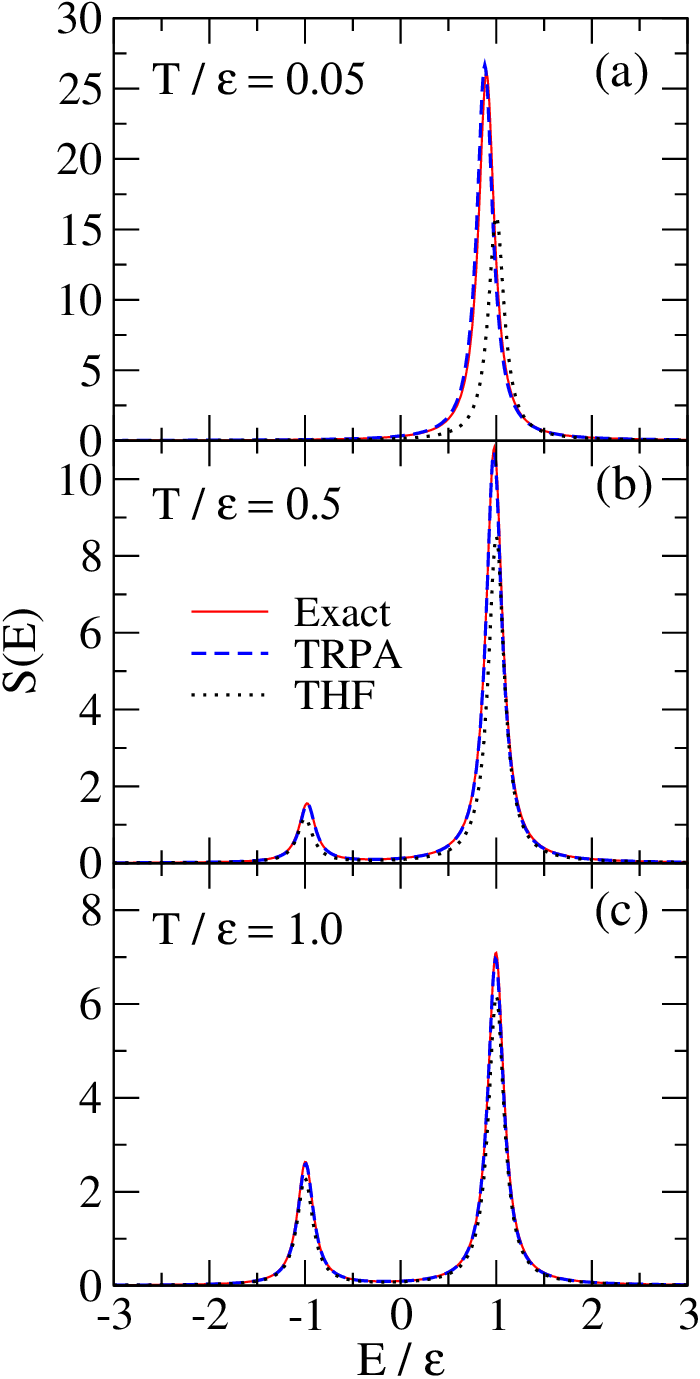}
\caption{(Color online)
The strength function for the operator 
$\hat{F}=(\hat{K}_++\hat{K}_-)/2$ 
obtained with the several methods for $VN/\epsilon$=0.5. 
Figs. 1(a), 1(b), and 1(c) correspond to the temperature of 
$T/\epsilon=0.05$, 0.5, and 1.0, respectively. 
The solid line is the exact result for the grand canonical ensemble, while the 
dashed and the dotted 
lines denote the solutions of finite temperature RPA and thermal Hartree-Fock, respectively. 
The exact result for the canonical ensemble is almost the same as the solid line, and is not shown in the figure. 
}
\end{figure}

\begin{figure}[htb]
\includegraphics[clip,scale=0.5]{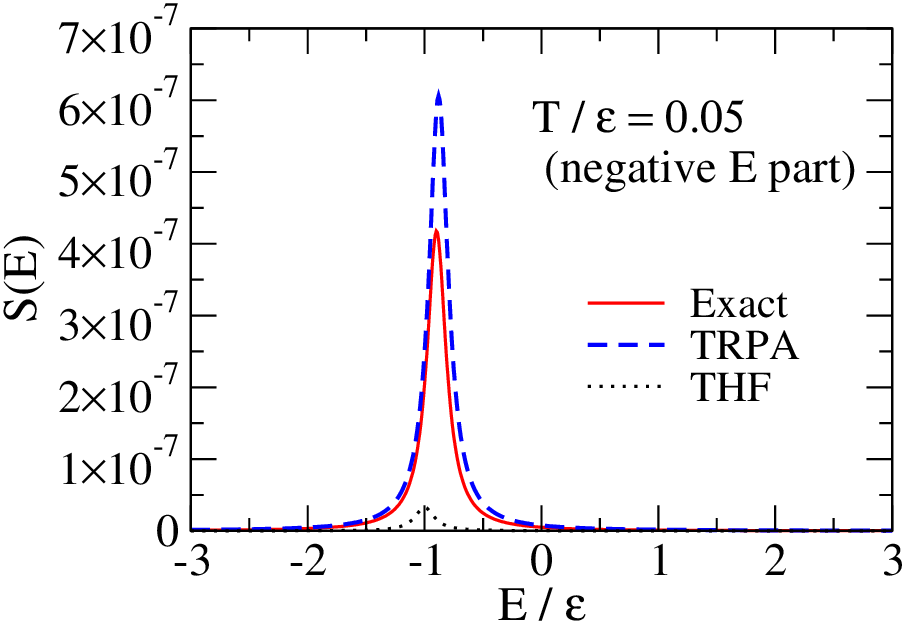}
\caption{(Color online) 
Same as Fig. 1(a), but only with the negative energy part. 
}
\end{figure}

At finite temperatures, the excited levels are thermally occupied with 
a finite probability. 
The probability for 
de-excitation of the excited states with the operator $\hat{F}$ 
appears in the negative 
energy part of the strength function. Fig. 2 shows the negative energy 
part of Fig. 1 (a). 
Since the temperature is low, the thermal occupation probability of the 
excited states is 
negligibly small. Nevertheless, we find that the RPA works reasonably well. 
The energy of the first excited state is 0.897$\epsilon$ for the exact 
solution, while 
it is 0.88$\epsilon$ in RPA.  As the energy is slightly underestimated 
in RPA, the peak 
of the strength function is somewhat overestimated. 
Despite this, we will show later that the temperature dependence of 
total strength 
is well reproduced with finite temperature RPA (see Figs. 3 and 4). 

The strength functions at higher temperatures, $T/\epsilon$=0.5 and 1.0, 
are shown in Figs. 1(b) and 1(c). At these temperatures also, one sees 
that the finite 
temperature RPA 
works well both for the positive and the negative energy parts. 
Especially, the shift of 
the peak position in the strength function due to the finite temperature 
effects is well 
reproduced with RPA. 
As the temperature increases, even the thermal Hartree-Fock method 
reproduces the exact 
strength function. 

The total strength defined as 
\begin{equation}
S_{\rm tot}=\int^\infty_{-\infty}S(E)\,dE,
\end{equation}
is plotted in the upper panel of Fig. 3 as a function of temperature. 
The lower panel shows the ratio of the total strength to 
that at zero temperature. 
The result of finite temperature RPA closely follows 
the exact result of grand canonical ensemble, 
as has been noted in Ref. \cite{RR98}. 
We have confirmed that this conclusion remains qualitatively the same 
even for a smaller number of particle number, {\it e.g.,} $N=10$. 
The result of canonical ensemble, 
shown by the thin solid line, 
is close to the result of grand canonical ensemble, although the difference is 
not negligible. 
The thermal Hartree-Fock, on the other hand, leads to an inconsistent 
temperature 
dependence of 
total strength, as can be seen in the lower panel. 

\begin{figure}[htb]
\includegraphics[clip,scale=0.4]{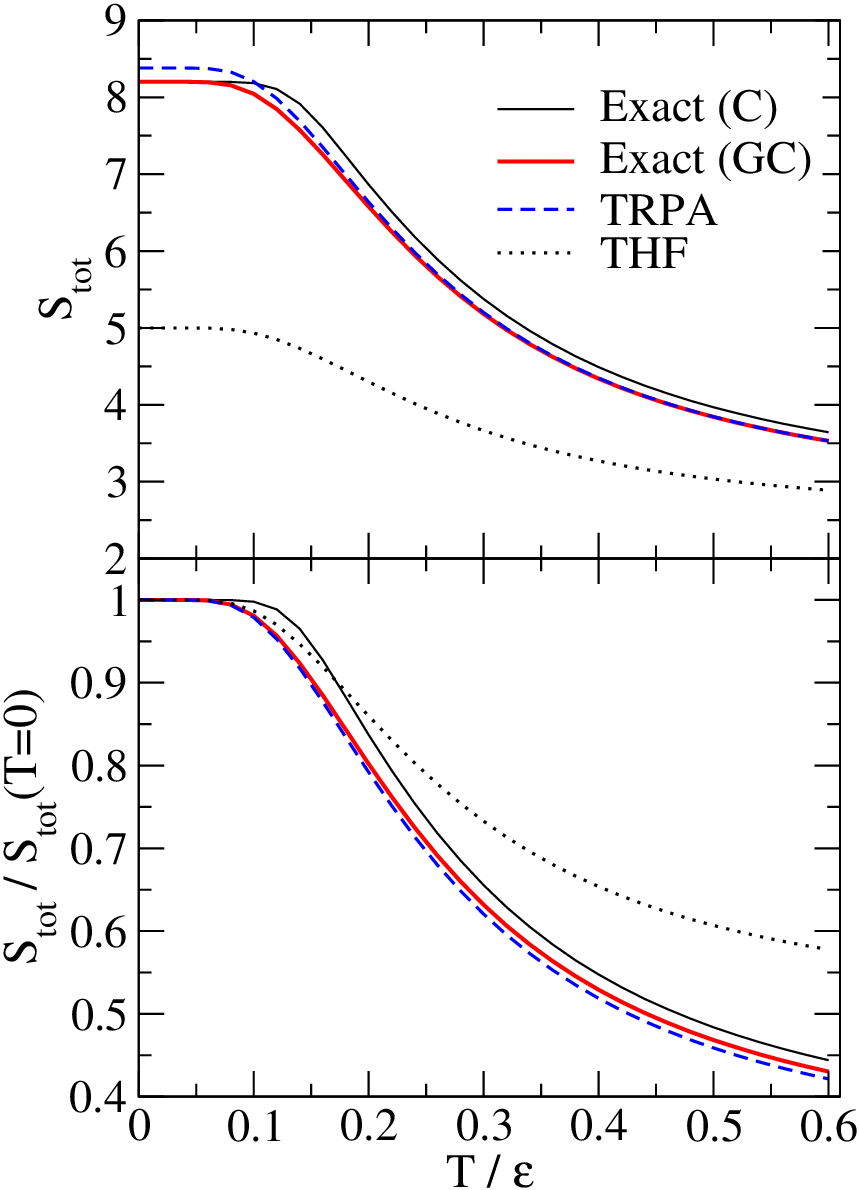}
\caption{(Color online)
The total strength obtained with 
the several methods as a function of temperature 
(the upper panel). 
The meaning of each line is the same as in Fig. 1, except for the thin 
solid line which 
denotes the exact result of canonical ensemble. The lower panel shows 
the ratio of the total 
strength to that at zero temperature. 
}
\end{figure}

\begin{figure}[htb]
\includegraphics[clip,scale=0.45]{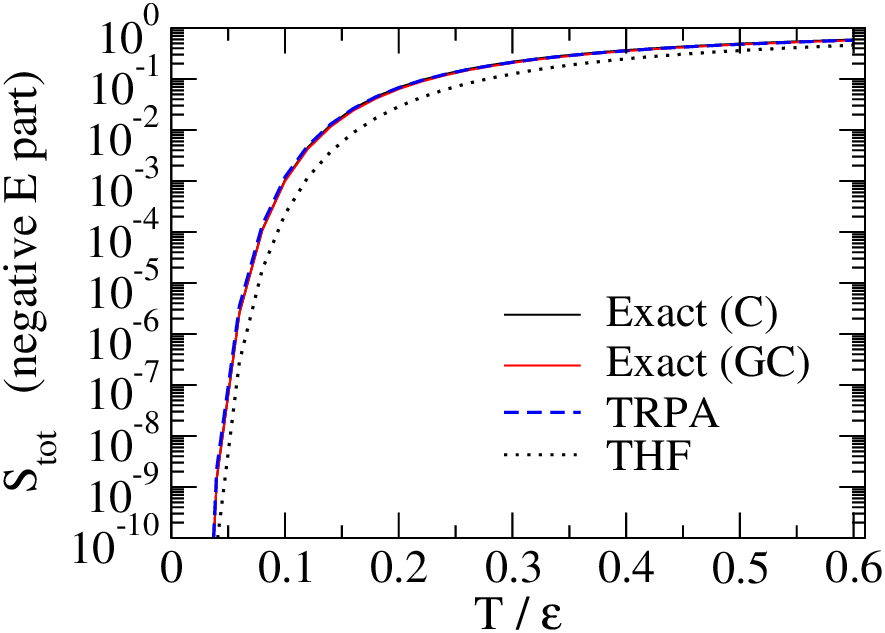}
\caption{(Color online)
Contribution of the negative energy part to the total strength shown in Fig. 3. 
The thin and the thick solid lines are indistinguishable on this scale. 
}
\end{figure}

The contribution of the negative energy part to the total strength is 
shown separately in Fig. 4. As one can see, the finite temperature RPA 
yields the correct 
temperature dependence of total strength even at very low temperatures. 

\begin{figure}[htb]
\includegraphics[clip,scale=0.5]{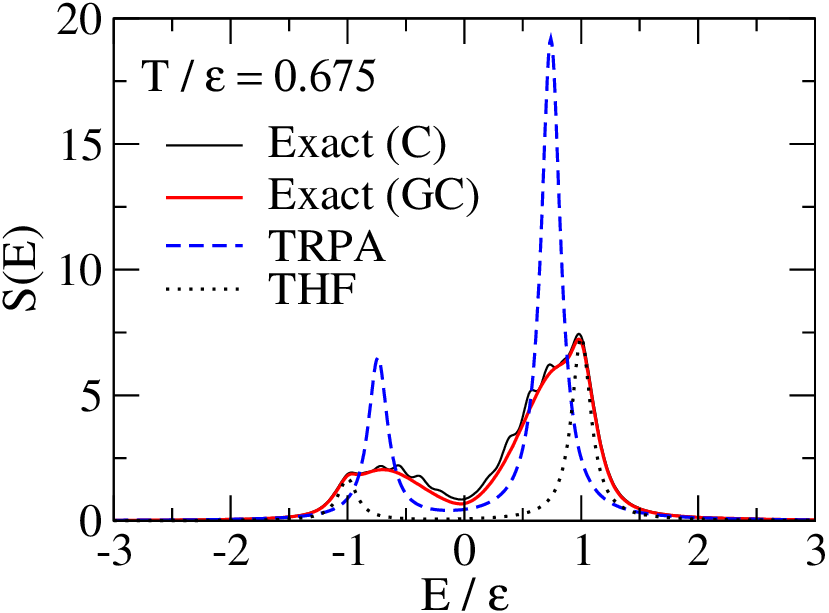}
\caption{(Color online)
Same as Fig.1, but for a stronger coupling strength, $VN/\epsilon$=2.0 
at temperature $T/\epsilon$=0.675. 
}
\end{figure}

Let us next discuss briefly a case with a stronger coupling. 
Figure 5 shows the strength function for 
$VN/\epsilon$=2.0 at temperature $T/\epsilon$=0.675. 
This corresponds to Fig. 2(b) in Ref. \cite{RR98}. 
In this case, the canonical and the grand canonical ensembles yield 
slightly different strength distributions from each other. That is, 
the grand canonical ensemble leads to a smoother strength function because of 
number fluctuation, although the overall behavior is similar to each other. 
One also notices that the finite temperature RPA significantly 
underestimates the thermal broadening 
of the strength function a/o the ground state correlation. 
However, as has been  argued in Ref. \cite{RR98}, 
the finite temperature RPA provides a reasonable estimate for the 
total strength. 
Table I summarizes the total strength 
for this particular choice for the parameters. 
The agreement between the exact results and 
the finite temperature RPA is satisfactory. 
We have checked that this is the case 
even for a stronger coupling, 
$VN/\epsilon$=4.0.  

\begin{table}[hbt]
\caption{
The total strength 
for $VN/\epsilon$=2.0 at temperature $T/\epsilon$=0.675 obtained 
with several methods, that is, 
the 
exact result with canonical (C) and 
grand canonical (GC) ensembles, thermal RPA (TRPA), and 
thermal Hartree-Fock (THF). 
The contribution from the negative energy 
part is also listed in the parentheses. 
}
\begin{center}
\begin{tabular}{cccc}
\hline
\hline
C & GC & TRPA & THF \\
\hline
7.92 & 7.40 & 8.04 & 2.81 \\
(2.02) & (1.84) & (2.01) & (0.52) \\
\hline
\hline
\end{tabular}
\end{center}
\end{table}

In summary, we have investigated the applicability of finite temperature 
RPA using a schematic solvable model. 
We have shown that the finite temperature RPA provides a reasonable 
estimate for the total strength, both for the excitation and the 
decay processes. This is the case even at low temperatures. 
For a small coupling case, the finite temperature RPA also yields 
a reasonable strength function itself. 
We have also shown that the canonical and the grand canonical ensembles 
lead to similar strength functions, as well as the total strengths, 
to each other. 
We thus conclude that the finite temperature RPA, being based on the 
grand canonical ensemble, provides a reasonable tool to discuss properties 
of hot and warm nuclei, including those in a stellar environment. 

\medskip

We thank G. Col\`o for useful discussions. 
This work was supported by the Japanese
Ministry of Education, Culture, Sports, Science and Technology
by Grant-in-Aid for Scientific Research under
the program number 19740115.


\begin{thebibliography}{99}

\bibitem{RS80}
P. Ring and P. Schuck, {\it The Nuclear Many Body Problem}
(Springer-Verlag, New York, 1980).

\bibitem{LG76}K.F. Liu and Nguyen Van Giai,
Phys. Lett. {\bf 65B}, 23 (1976). 

\bibitem{GK02}S. Goriely and E. Khan, 
Nucl. Phys. {\bf A706}, 217 (2002). 

\bibitem{B55}D.M. Brink, Ph.D. thesis, Oxford University, 1955 (unpublished). 

\bibitem{N81}J.O. Newton {\it et al.}, Phys. Rev. Lett. {\bf 46}, 
1383 (1981). 

\bibitem{B89}A. Bracco {\it et al.}, Phys. Rev. Lett. {\bf 62}, 2080 (1989); 
Phys. Rev. Lett. {\bf 74}, 3748 (1995). 


\bibitem{S86}K.A. Snover, Ann. Rev. Nucl. Part. Sci. {\bf 36}, 545 (1986). 

\bibitem{Y90}K. Yoshida {\it et al.}, Phys. Lett. {\bf B245}, 7 (1990). 

\bibitem{K99}M.P. Kelly {\it et al.}, Phys. Rev. Lett. {\bf 82}, 3404 (1999). 


\bibitem{VVM84}D. Vautherin and N. Vinh Mau, Phys. Lett. {\bf 120B}, 
261 (1983);  Nucl. Phys. {\bf A422}, 140 (1984). 

\bibitem{FER83}M.E. Faber, J.L. Egido, and P. Ring, Phys. Lett. {\bf 127B}, 
5 (1983). 

\bibitem{RREF84}P. Ring {\it et al.}, 
Nucl. Phys. {\bf A419}, 261 (1984). 

\bibitem{ER93}J.L. Egido and P. Ring, J. of Phys. {\bf G19}, 1 (1993). 

\bibitem{Som83}H.M. Sommermann, Ann. of Phys. {\bf 151}, 163 (1983). 

\bibitem{SB84}H. Sagawa and G.F. Bertsch, Phys. Lett. {\bf 146B}, 
138 (1984). 

\bibitem{BRT84}W. Besold, P.-G. Reinhard, and C. Toepffer, 
Nucl. Phys. {\bf A431}, 1 (1984). 

\bibitem{CBD84}O. Civitarese, R.A. Broglia, and C.H. Dasso, 
Ann. of Phys. {\bf 156}, 142 (1984). 

\bibitem{RBR99}C. Rei\ss, M. Bender, and P.-G. Reinhard, 
Eur. Phys. J. {\bf A6}, 157. 

\bibitem{LKD01}K. Langanke, E. Kolbe, and D.J. Dean, 
Phys. Rev. C{\bf 63}, 032801(R) (2001). 

\bibitem{NK04}J.-U. Nabi and H.V. Klapdor-Kleingrothaus, 
At. Data Nucl. Data Tables {\bf 88}, 237 (2004). 

\bibitem{CHMR00}O. Civitarese {\it et al.}, 
Phys. Rev. C{\bf 62}, 054318 (2000). 

\bibitem{NPVM09}Y.F. Niu {\it et al.}, 
arXiv:0906.2973 [nucl-th].

\bibitem{KGG04}E. Khan, Nguyen Van Giai, and M. Grasso, Nucl. Phys. 
{\bf A731}, 311 (2004). 

\bibitem{CR99}O. Civitarese and A. Ray, Phys. Scr. {\bf 59}, 352 (1999). 

\bibitem{LMG65}H. Lipkin, N. Meshkov, and A.J. Glick, 
Nucl. Phys. {\bf 6}, 188 (1965). 

\bibitem{RR98}R. Rossignoli and P. Ring, Nucl. Phys. A{\bf 633}, 613 (1998). 

\bibitem{TEKO94}
S.Y. Tsay Tzeng {\it et al.}, 
Nucl. Phys. {\bf A580}, 277 (1994). 

\bibitem{RPM91}R. Rossignoli, A. Plastino, and H.G. Miller, 
Phys. Rev. C{\bf 43}, 1599 (1991). 

\bibitem{CVVM90}Ph. Chomaz, D. Vautherin, and N. Vinh Mau, 
Phys. Lett. {\bf B242}, 313 (1990). 




\end{thebibliography}
\end{document}